\documentclass[11pt, letter]{article}

\usepackage{stolenstyle}

\usepackage{amsmath,epsf,amssymb,latexsym,amsthm,setspace,bbm,array,pifont}

\usepackage[matrix,arrow,frame,import,curve,color]{xy}

\DeclareFontFamily{U}{rsf}{}
\DeclareFontShape{U}{rsf}{m}{n}{
  <5> <6> rsfs5 <7> <8> <9> rsfs7 <10-> rsfs10}{}
\DeclareMathAlphabet\Scr{U}{rsf}{m}{n}

\def\CO#1#2{{[#1,#2]}}
\def\AC#1#2{{\{#1,#2\}}}

\def\iden{{\mathbbm 1}}

\def\rep#1{{{\boldsymbol{#1}}}}

\def\R{{\mathbb R}}
\def\Z{{\mathbb Z}}

\def\Img{\operatorname{Im}}

\def\Rea{\operatorname{Re}}

\def\Tr{\operatorname{Tr}}

\def\dVol{\operatorname{dVol}}

\def\diag{\operatorname{diag}}

\def\tr{\operatorname{tr}}

\def\SO{\operatorname{SO}}

\def\GO{\operatorname{O{}}}
\def\SU{\operatorname{SU}}
\def\GU{\operatorname{U{}}}
\def\Sp{\operatorname{Sp}}
\def\Spin{\operatorname{Spin}}
\def\GG{\operatorname{G}}

\def\GE{\operatorname{E}}

\def\p{\partial}

\def\ff#1#2{{\textstyle\frac{#1}{#2}}}

\def\cE{{\cal E}}
\def\cF{{\cal F}}

\def\cJ{{\cal J}}

\def\cL{{\cal L}}

\def\cN{{\cal N}}

\def\cP{{\cal P}}

\def\cX{{\cal X}}

\def\ep{{\epsilon}}

\newcommand\xit{\widetilde{\xi}}

\newcommand\omegat{\widetilde{\omega}}

\newcommand\vphi{\varphi}

\newcommand\Thetat{\widetilde{\Theta}}

\newcommand\et{\widetilde{e}}

\newcommand\jt{\widetilde{\jmath}}

\newcommand\mt{\widetilde{m}}

\newcommand\ttld{\widetilde{t}}  

\newcommand\vt{\widetilde{v}}

\newcommand\xt{\widetilde{x}}

\newcommand\Ah{\widehat{A}}

\newcommand\Et{\widetilde{E}}

\newcommand\Rt{\widetilde{R}}
\newcommand\St{\widetilde{S}}

\newcommand\Vt{\widetilde{V}}

\newcommand\Xt{\widetilde{X}}

\def\bea#1\eea{\begin{align}#1\end{align}}

\def\bes #1\ees{\begin{split}#1\end{split}}

\newcommand{\be}{\begin{equation}}
\newcommand{\ee}{\end{equation}}

\def\bF{{\boldsymbol{F}}}

\def\bV{{{\boldsymbol{V}}}}
\def\bF{{{\boldsymbol{F}}}}

\def\cXt{{\widetilde{\cX}}}

\def\preprint{  {{EFI-17-3}}\\ {{IPhT-T17/025}}  }

\title{Non--duality in three dimensions}
\author[a] {Ilarion V.~Melnikov,}
\author[b] {Ruben Minasian,}
\author[\,c] {and Savdeep Sethi}
\affiliation[a]{Department of Physics and Astronomy, James Madison University, VA 22807, USA}
\affiliation[b]{Institut de Physique Th{\'e}orique, Universit{\'e} Paris Saclay, CNRS, CEA \\ Orme des Merisiers, F-91191 Gif-sur-Yvette, France}
\affiliation[c]{Enrico Fermi Institute, University of Chicago, Chicago, IL 60637, USA}
\emailAdd{melnikix@jmu.edu}
\emailAdd{ruben.minasian@cea.fr}
\emailAdd{sethi@uchicago.edu}
\abstract{We investigate M-theory and heterotic compactifications to 7 and 3 dimensions.  In 7 dimensions we discuss a class of massive supergravities that arise from M-theory on K3 and point out obstructions to realizing these theories in a dual heterotic framework with a geometric description.  Taking M-theory further down to 3 dimensions on $\text{K3}\times\text{K3}$ with a choice of flux leads to a rich landscape of theories with various amounts of supersymmetry, including those preserving 6 supercharges.   We explore possible heterotic realizations of these vacua and prove a no--go theorem:  every heterotic geometry that preserves 6 supercharges preserves 8 supercharges.  }

\begin{document}

\maketitle

\section{Introduction} \label{s:intro}

Ever since its discovery~\cite{Witten:1995ex}, the seven-dimensional duality between M-theory compactified on K3 and the heterotic string compactified on $T^3$ has played an essential role in our understanding of string theory.  In its most prosaic usage, this duality, together with its F-theory counterpart~\cite{Vafa:1996xn}, have been used to produce and study many dual pairs of theories.  The simplest way to generate such pairs is to compactify the resulting seven-dimensional theory on some common geometry; however, a much richer class of theories can be obtained by performing a fiber-wise duality.  A classic example of this correspondence is between M-theory on $\text{K3}\times\text{K3}$ and heterotic flux vacua with target space a principal $T^3$ bundle over K3~\cite{Dasgupta:1999ss}.  Depending on the choice of $G$-flux on the M-theory side, these may have an F-theory lift with a corresponding heterotic vacuum that is a principal $T^2$ bundle over K3.  The latter have been the focus of much attention, e.g.~\cite{Becker:2006et,Becker:2009df,Melnikov:2012cv,Melnikov:2014ywa}, and remain the sole compact examples of heterotic flux vacua.

The class of four- or three-dimensional vacua just discussed is particularly simple, and with a sufficient amount of supersymmetry, it should be possible to work out the duality map in some detail.  However, these solutions are by no means exhaustive.  Surprisingly, there are many M-theory vacua on $\text{K3}\times\text{K3}$ that are more challenging to  describe from a heterotic perspective.  The main aims of this work are to present these vacua, describe their basic features, and to point out the challenges in finding corresponding heterotic descriptions.

Given that the M-theory geometry $M_8 = \text{K3}\times\text{K3}$ is so simple, the reader will not be surprised that it is the choice of $G$-flux that is responsible for the extra complications.  It is possible, while preserving $\cN=1,2,3$ three-dimensional super-Poincar{\'e} invariance\footnote{The $\cN$ counts the two-component Majorana spinors of $\R^{1,2}$.}, to choose $G$ with a component that threads the volumes of the two K3 factors.  Such a volume-threading flux automatically obstructs a lift to M-theory on $\text{K3}\times\R^{1,6}$, so that we cannot apply a standard duality with heterotic string on $T^3 \times \R^{1,6}$.  

On the other hand, given that we do have the $3d$ solutions with this sort of volume-threading flux, it is clear that there exists a supergravity theory in seven dimensions obtained by reducing M-theory on K3 surface $X$ with a volume-threading flux $G \supset \text{const} \times \dVol(X)$ that has supersymmetric vacua of the form $\R^{1,2}\times \Xt$, where $\Xt$ is a second K3.  This seven-dimensional theory does not have Minkowski vacua, and, as we will argue, necessarily involves a spacetime potential and cosmological constant.  The volume-threading flux will obstruct any lift to F-theory, and it will lead to a puzzle with any potential geometric heterotic dual:  briefly stated, the Bianchi identity for the $H$-flux of the putative seven-dimensional heterotic dual theory would seem to involve a term of the form $d H =  \ast H + \cdots$, where $\cdots$ refer to the familiar heterotic curvature terms; the $\ast H$ term is not present in standard formulations of the heterotic string. 

This is a strong indication that there is no conventional dual description of M-theory on K3 with a volume-threading flux.   We sharpen this statement as follows.   First, we demonstrate in some detail that in M-theory on $X\times \Xt$ it is possible to have solutions that preserve three-dimensional $\cN=3$ super-Poincar{\'e} invariance.  These sorts of vacua are interesting in themselves, since they lie between the challenging $\cN=2$ and the reasonably well-understood $\cN=4$ vacua.  We show that these solutions necessarily involve volume-threading flux, and this is why they have not been previously encountered in the literature (for instance, they certainly do not have an F-theory lift).   We next turn to the heterotic string, and we prove that every geometric compactification with $\cN=3$ invariance in fact preserves $\cN=4$.\footnote{By a geometric compactification we mean a solution (perhaps with some formal $\alpha'$ expansion~\cite{Melnikov:2014ywa} with a smooth seven-dimensional geometry $X_7$ equipped with some gauge bundle, dilaton, and $H$-flux.}

This work should be viewed as an exploration of the general structure of duality between M-theory and the heterotic string.  Consider a compactification of M-theory based on an $8$-dimensional Ricci-flat manifold $M_8$.  Compactification geometries with extended supersymmetry are conveniently summarized by the following famous table.\footnote{Like Kodaira's list of singularities, the M-theory part of the table can be found in many string theory papers; we adapt it from~\cite{Joyce:2000cm}.} The maximum number of supersymmetries $\cN_{\text{max}}=\Ah(M_8)$, where  $\Ah(M_8)$ is the value of the Dirac index on $M_8$.
\begin{center}
\begin{tabular}  {c| c c c c c}
M-theory		& $\cN_{\text{max}}$ 	&	1			&2				&3				&4 \\[2mm]
~			& $M_8$ holonomy	&$\Spin(7) \supset$	&$\SU(4)\supset$	&$\Sp(2) \supset$	&$\Sp(1)\times\Sp(1)$ \\[2mm] \hline
Heterotic 		& ~				&$\cup$			&$\cup$			&~~		&\rotatebox{90}{$\subseteq$} \\[2mm] 
~			&$X_7$ structure	&$G_2 \supset$	&$\SU(3) \supset$	&???& 	$ \Sp(1)$ 
\end{tabular}
\end{center}
The lower line of the table  can also be taken to  indicate the holonomy of an internal eight-manifold in M-theory compactifications $M_8$, rather than the structure group of $X_7$ on  the heterotic side. For all of these the $8d$ spinors are not chiral,  and  $\Ah(M_8)=0$. Moreover, the internal flux $G$ vanishes. The amount of $3d$ supersymmetry changes as $\cN=2,4,8$ as one moves right along the line. $\cN=16$ corresponds to $M_8$ with trivial holonomy. All these theories have natural lifts to four dimensions, since $M_8$ will necessarily involve at least one trivial circle.  In this work we will not consider such flux-free M-theory compactifications.

The table, while specifying the geometry, does not describe the conditions on the flux of M-theory or the choice of gauge bundle of heterotic compactifications.  For each class of $M_8$ it may be possible to choose $G$ to preserve the maximal $\cN_{\text{max}}$ supersymmetry; we know many examples of this form, and for a very large class of solutions (in particular those with an F-theory lift), we have conjectured (and in examples tested) dual pairs of M-theory/heterotic theories.     

The solutions with $\cN=3$ are certainly the least familiar in the list, and we conclude our introduction by making two general points about them.  First, we have a rather poor understanding of M-theory vacua based on $M_8$ with $\Sp(2)$ holonomy.  This is in part due to a dearth of examples of hyper-K\"ahler manifolds; a primary example is a resolution of the symmetric product, $S^2(K3)$, of two K3 surfaces.\footnote{An analysis of flux choices for this case, using orbifold techniques, can be found in~\cite{Dasgupta:1999ss}.} However, as our heterotic no--go result shows, as far as duality goes, the issue appears to be deeper:  there are no candidates for dual heterotic geometries.  Second, although  general $\Sp(2)$ manifolds may be of our reach, there does not appear to be much of a difference from the perspective of the $\R^{1,2}$ spacetime theory between $M_8$ with $\Sp(2)$ holonomy or $M_8 = X\times\Xt$ with an appropriate choice of flux; so, even with existing geometric technology there are many $\cN=3$ vacua to be explored.  We will discuss their 
most basic properties below.
\vspace{1cm}

The rest of this paper is organized according to the table of contents. Seven-dimensional dualities are discussed in section \ref{s:M7}. Sections \ref{s:MonK3K3} and \ref{s:heterotic} are devoted to M-theory on $\text{K3}\times\text{K3}$ and  heterotic three-dimensional compactifications respectively.  An appendix contains some technical details.

\section{M--theory in seven dimensions} \label{s:M7}

Let $X$ be a K3 equipped with a hyper-K\"ahler metric.  We denote the triplet of hyper-K\"ahler forms by $j_a$, $a=1,2, 3$, and we normalize them by
\begin{align}
j_a \wedge j_b = 2\delta_{ab} v E~,
\end{align}
where $v$ is the volume of $X$ and $E$ is the generator of $H^4(X,\Z)$.  In addition, we have the $19$ anti-self-dual forms $\omega_\alpha$, $\alpha=1,\ldots, 19$ that satisfy
\begin{align}
\omega_\alpha \wedge j_a & = 0~, &
\omega_\alpha \wedge \omega_\beta & = -2\delta_{\alpha\beta} v E~.
\end{align}
In what follows we will often suppress the explicit $\wedge$ when there is no possibility of confusion.

These conditions are invariant with respect to $\SO(3)\times\SO(19)$ rotations that act on the $j_a$ and $\omega_\alpha$ in the obvious fashion.   The triplet $j_a$ defines an $\SU(2)$ structure; in particular, the $j_a$ determine the metric in the following way:  a combination of two of them, say $j_2 + i j_3$, determines an integrable complex structure, and then the orthogonal complement, in this case $j_1$, becomes a corresponding K\"ahler form.  There are $\SO(3)/\GU(1) = S^2$ ways of picking a complex structure and, evidently, every $\SO(3)$ rotation of $j_a$ yields exactly the same Einstein metric.  The double cover $\SU(2)$ of this $\SO(3)$ turns out to be the $\SU(2)$ R-symmetry of the $7$--dimensional theory.

\subsection{Dualities between massive theories}

We are interested in the physics of M-theory compactified on $X$ with volume $v$. In the absence of any flux, this background is famously dual to the heterotic or type I string compactified on $T^3$. This is a strong-weak duality with 
\begin{align}
e^{\phi_7} = v^{3/4}, 
\end{align}
where $e^{\phi_7}$ is the $7$-dimensional heterotic string coupling. For elliptic $X$ with section, this background has an $8$-dimensional F-theory limit, corresponding to decompactifying a circle of the heterotic $T^3$.  

We would like to ask whether a volume threading flux, $G \supset \text{const} \times \dVol(X)$, which is compatible with Lorentz invariance in $7$ dimensions, admits a dual description. Kaluza-Klein reduction on this background was first studied in~\cite{Lavrinenko:1996mp, Lu:1996rh}. At first sight, the question itself might appear strange. The background with flux is not a solution of the equations of motion with an $\R^{1,6}$ Minkowski spacetime. This is easy to see for an unwarped spacetime metric from the M-theory equations of motion,  
\be
{\cal R}_{\mu\nu} = -{1\over 6}\eta_{\mu\nu} |G|^2,  
\ee
since the spacetime Ricci tensor, which should vanish, is sourced by the flux. For a warped background, we consider the metric Ansatz
\be
ds^2 = e^{2\omega} \eta + ds^2_{X}~,
\ee
where $\eta$ denotes the usual Minkowski metric, and $\omega$ is the warp factor. For warp factors that do not depend on the spacetime coordinates, the Ricci tensor takes the form
\be
{\cal R}_{\mu\nu} = - {1\over 5} \eta_{\mu\nu} e^{-5\omega} \nabla^2 e^{5\omega}. 
\ee
However on a compact space like $X$, the equation
\be
\nabla^2 e^{5\omega} = {5\over 6} e^{5\omega} |G|^2
\ee
has no solution. Therefore there are no Minkowski vacua for M-theory reduced on $X$ with volume threading $G$. Indeed, it is easy to extend this argument to see that there are no solutions for any maximally symmetric spacetime:  this background cannot be realized as an on-shell solution in string theory without breaking maximal spacetime symmetry. Alternately, it can appear as an intermediate massive theory en route to a static Minkowski or AdS solution in lower dimension. 

Regardless, we can still perform a Kaluza-Klein reduction on such a background and some theory must determine the set of quantum corrections to the classical spacetime effective action.
From this latter perspective, it is still reasonable to ask whether a weakly coupled description might control the physics of small volume $v$, while eleven-dimensional supergravity together with higher derivative corrections controls the perturbative physics of large $v$.  A natural guess based on the flux-free duality might be that the heterotic string on $T^3$ with a volume threading flux $H  \supset \text{const} \times \dVol(T^3)$ provides such a description.  The gauged massive supergravities that arise from toroidal compactifications of the heterotic string with backgrounds fluxes have been studied in~\cite{Kaloper:1999yr}. In both cases, decompactification to $8$ dimensions is obstructed by the quantized flux.

However, there are immediate issues with such a proposed duality. To construct a macroscopic heterotic string, we usually wrap an M5-brane on $X$.  The M5-brane world-volume supports a self-dual $3$-form field strength $h_3$, which obeys a Bianchi identity:
\be
dh_3 =G. 
\ee
This obstructs wrapping $X$ without some added ingredient to satisfy the Gauss law charge constraint; for example, stretched M2-branes which realize self-dual strings on the world-volume of the M5-brane. However, any such ingredient breaks additional Lorentz invariance beyond the breaking introduced by the stretched macroscopic string. 

Another immediate issue is seen by examining the form of the Bianchi identity for the heterotic $H$-flux, derived from M-theory. In the flux-free duality, we identify
\be
H = \ast_7 G, 
\ee  
where the Hodge dual is taken in $7$ dimensions; the Bianchi identity for $H$ follows from the equation of motion for $G$. Since these are the only propagating $3$-forms in $7$ dimensions, any proposed duality would need some identification between them. However the case with flux threading $X$ produces a new coupling in the heterotic Bianchi identity:
\be\label{hetbianchi}
\int_X d\ast G = -{1\over 2} \int_X G \wedge G \quad \implies \quad dH \sim \ast_7 H + \ldots. 
\ee
The omitted terms involve both gauge-fields from the Kaluza-Klein reduction of $G$ on $2$-forms of $X$, and gravitational couplings from higher derivative interactions in M-theory. The new coupling involving $\ast_7 H$ in~(\ref{hetbianchi}), which is proportional to the amount of flux through $X$, has no obvious realization in geometric heterotic compactifications.  

On its own, these pieces of evidence might not be convincing. It might be the case that heterotic compactified on $T^3$ with $H$-flux simply admits no macroscopic string solutions, and perhaps there is some subtle modification of the Bianchi identity to evade~(\ref{hetbianchi}). There is a more direct approach.
To dualize M-theory on $X$ to a type I/heterotic background,   start with an orbifold limit where $X=T^4/\Z_2$. The first step in the duality chain is to reduce on a circle of $X$ to a type IIA orientifold: 
\be
T^3/\Omega\Z_2. 
\ee
The $G$-flux through $X$ implies a volume threading $H$-flux in this type IIA background. Arriving at a type I background requires T-dualizing all three directions of the $T^3$. While one or two T-dualities can be performed in this background without great difficulty, dualizing all three directions is difficult to understand. Each T-duality requires a potential for $H$, but any trivialization of $H$ breaks one isometry of $T^3$. For more discussion of what such a resulting heterotic theory might possibly look like, see, for example,~\cite{Blumenhagen:2011ph}. This obstruction looks difficult to evade, and each known duality chain that leads to a heterotic or type I dual description meets this same issue in one guise or the other. 

On the other hand, heterotic on $T^3$ with volume threading $H$-flux does admit a dual description, which can be seen as follows: consider heterotic on $T^2$. There are $2$ periodic scalars $(\tau_1, \rho_1)$ associated to the complex structure of $T^2$ and to the volume threading $B_2$-flux. Let us focus on this latter scalar. Under the duality to the type IIB orientifold $T^2/\Omega(-1)^{F_L}\Z_2$ described in~\cite{Sen:1997gv}, this scalar maps to the type IIB axion $C_0$. Now compactify on a further $S^1$ and permit both $\rho_1$ of heterotic and $C_0$ of type IIB to depend linearly on the circle coordinate. On the one side, we have heterotic on $T^3$ with volume threading $H$. The dual description is type IIB compactified on
\be
T^2/\Omega (-1)^{F_L}\Z_2 \times S^1 
\ee
with constant quantized RR $F_1$ field strength in the $S^1$ direction.  This is a dual pair. The usual route to find a lift to M-theory involves T-dualizing on the $S^1$ direction. This maps $F_1$ to $F_0$ and we arrive at a massive type IIA background~\cite{Romans:1985tz}.\footnote{ The interpretation of massive theories from the perspective of M-theory and F-theory has been described in~\cite{Hull:1998vy}.} The Romans mass obstructs a lift to M-theory.  
The dual description is actually given by either the type IIA or type IIB $7$-dimensional orientifold background, depending on the size of $S^1$. This story can be extended to a more general F-theory setting by identifying both periodic scalars $(\tau_1, \rho_1)$ in the geometry of an elliptic $X$~\cite{McOrist:2010jw, Garcia-Etxebarria:2016ibz}, and allowing them to depend on the $S^1$ coordinate, along the lines described in~\cite{Cowdall:1996tw}.

The bigger picture suggested by this example is a collection of dualities between lower-dimensional massive supergravity theories, induced from higher-dimensional more conventional dual pairs. Interestingly, there is no obvious candidate for a dual description of M-theory on $X$ with volume threading $G$. Like the Romans theory in ten dimensions, it might simply exist without a relation to other known massive backgrounds. 


\section{M--theory vacua on $\text{K3}\times\text{K3}$} \label{s:MonK3K3}

In the previous section we discussed M--theory compactifications to seven dimensions with a $G$--flux threading the K3.  We saw that such theories do not have $\R^{1,6}$ vacua; on the other hand, it is possible to compactify further and obtain $\R^{1,2}$ vacua.  In this section we explore the resulting theories in some detail.   We will find familiar examples of three--dimensional $\cN = 1$, $\cN=2$ and $\cN=4$ supersymmetric theories, but we will also find solutions that are a bit more exotic and realize $\cN=3$.

The interest in these solutions is two--fold.  First, we will be able sharpen the puzzles of ``non-duality,'' because, as we will see in the section that follows, there are no geometric $\cN=3$ heterotic compactifications.  Second, we will see that the choice of flux allows the rather simple geometry of K3$\times$K3 to realize the features of more sophisticated solutions with $\Spin(7)$, $\SU(4)$ or $\Sp(2)$ structures. 

\subsection{M--theory on $M_8$}
We begin with a brief review of M--theory compactification of the form $\R^{1,2}\times M_8$.  This is determined by a choice of a warped metric $g$ and flux $G$ on $M_8$.  The latter needs to obey two topological conditions: the flux is quantized according to~\cite{Witten:1996md}
\begin{align*}
\ff{1}{2\pi} G - \ff{1}{4} p_1(M_8) \in H^4(M_8,\Z)~,
\end{align*}
and it satisfies the tadpole equation for $C_3$~\cite{Duff:1995wd, Becker:1996gj}.  A necessary and sufficient condition for that is
\begin{align}
\frac{1}{2} \int_{M_8} \frac{G}{2\pi} \wedge \frac{G}{2\pi} = \frac{\chi(M_8)}{24} - N(M_2)~.
\end{align}
Here $\chi(M_8)$ is the Euler number of $M_8$, while $N(M_2)$ is the number of space-filling $M2$-branes.  In this paper we will be interested in solutions with $N(M_2) = 0$.

Minimal supersymmetry requires $M_8$ to admit a $\Spin(7)$ holonomy metric and also imposes a condition on $G$. $H^4(M_8,\R)$ can be decomposed according to representations of $\Spin(7)$~\cite{Joyce:2000cm}, and we must have $G \in H^4_{+\rep{27}}(M_8,\R)$.  That is, the flux is self-dual and in the $\rep{27}$~\cite{Acharya:2002vs}.

\subsection{M-theory on K3$\times$K3 and $\cN = 1,2,3,4$ examples}
We consider M-theory on $M_8 = X \times\Xt$, where $X$ and $\Xt$ are both K3 surfaces.  Minkowski $\R^{1,2}$ vacua with this compactification geometry are labeled by a choice of Einstein metric on $M_8$ and a choice of $G$ obeying the integrality and supersymmetry conditions.  In our case $p_1(M_8) = p_1(X) + p_1(\Xt)$ is  divisible by $4$, so the integrality condition on  $G$ is simply that $\ff{1}{2\pi} G \in H^4(X,\Z)$.  We have the identification 
\begin{align}
H^4(M_8,\Z) = H^2(X,\Z) \otimes H^2(\Xt,\Z) \oplus H^4(X,\Z) \oplus H^4(\Xt,\Z)~.
\end{align}
In much of the work on this compactification, e.g.~\cite{Dasgupta:1999ss,Aspinwall:2005qw}, the flux does not involve any components in the last two terms.  However, as has been observed more recently~\cite{Prins:2013wza}, minimal supersymmetry allows a more general flux. 

Consider now $X\times\Xt$.  The two components $X$ and $\Xt$ have self-dual forms $j_a$ and $\jt_{\dot a}$ and anti-self-dual forms $\omega_\alpha$, $\omegat_{\dot\alpha}$ in an obvious extension of the notation from section~\ref{s:M7}~.  We denote the generators of $H^4(X,\Z)$ and $H^4(\Xt,\Z)$ by, respectively, $E$ and $\Et$; similarly $X$ and $\Xt$ have volumes $v$ and $\vt$.   With this notation, we can state the general result~\cite{Prins:2013wza}:  up to an $\SO(3)\times\SO(3)$ rotation, the most general form of the G-flux on $X\times\Xt$ consistent with $\cN=1$ supersymmetry in $\R^{1,2}$ is
\begin{align}
\label{eq:PTflux}
G & = j_{a} M_{a \dot a} \jt_{\dot a} + (4A-2C) \left[ v E + \vt \Et\right] + f_{\alpha\dot\alpha} \omega_\alpha\omegat_{\dot\alpha}~,
\end{align}
and the $3\times 3$  constant matrix $M$ has the form
\begin{align}
M = \begin{pmatrix} C & D_1 & D_2 \\ D_1 & A+B_1 & B_2 \\ -D_2 & -B_2 & B_1 - A\end{pmatrix}~.
\end{align}
The last term in $G$ with the $19\times 19$ constant matrix $f$ just involves the anti-self-dual forms $\omega$ and $\omegat$.
$G$ should also satisfy the integrality and tadpole constraints.  

We can now give some examples of solutions that preserve different amounts of supersymmetry.
\begin{enumerate}
\item
The most familiar way to satisfy~(\ref{eq:PTflux}) is to set $M = 0$;  this also eliminates the ``volume--threading'' term.  In this case
$G$ is invariant under $\SO(3)\times\SO(3)$ rotations and corresponds to an $\cN = 4$ vacuum.  We can think of this as two statements:  the underlying manifold
has $\SU(2)\times\SU(2)$ structure, and the flux respects this:
\begin{align}
j_a  \wedge G & = 0~, &
\jt_{\dot a} \wedge G & = 0~.
\end{align}
\item  
We can reduce supersymmetry by taking $M = \lambda \iden_3$, so that 
\begin{align}
G & = \lambda (j_1 \jt_1 + j_2 \jt_2 + j_3 \jt_3) -2\lambda \left[ v E + \vt \Et\right] + f_{\alpha\dot\alpha} \omega_\alpha\omegat_{\dot\alpha}~.
\end{align}
$G$ is invariant under a diagonal $\SO(3)\subset \SO(3)\times\SO(3)$ action; in fact it respects an $\Sp(2)$ structure on the underlying manifold and therefore leads to $\cN=3$ in $\R^{1,2}$.  The $\Sp(2)$ structure is generated by the three $2$--forms\footnote{\label{foot:sp2} That is, the $\cJ_a$ are three non-degenerate two-forms that satisfy the defining cubic and quartic relations
$3\cJ_a\cJ_b \cJ_c  =  \delta_{ab} \cJ^3_c + \delta_{ca} \cJ^3_b + \delta_{bc} \cJ_a^3$ and
$\cJ_a \cJ_b \cJ_c \cJ_d  = 8 \dVol_8 \left[\delta_{ab} \delta_{cd} + \delta_{ca} \delta_{bd} + \delta_{bc} \delta_{ad} \right]$ \cite{Gauntlett:2003cy}.
}
\begin{align}
\cJ_a & = j_a + \jt_{a}~,
\end{align}
and for all $a$
\begin{align}
\cJ_a \wedge G & = 0~.
\end{align}

\item To obtain~$\cN=2$ symmetry we demand that $G$ only preserved by $\GU(1) \subset \SO(3) \subset \SO(3)\times\SO(3)$.  For instance,
following~\cite{Dasgupta:1999ss} we can take
\begin{align}
G & = \lambda ( j_1 \jt_1 + j_2  \jt_2) +f_{\alpha\dot\alpha} \omega_\alpha\omegat_{\dot\alpha}~.
\end{align}
This flux respects an $\SU(4)$ structure of $X\times\Xt$:  we set
\begin{align}
J &= \cJ_1~, & \Omega & = \ff{1}{2} (\cJ_2 + i\cJ_3)^2~,
\end{align}  
and $G$ is (2,2) and primitive with respect to this $\SU(4)$ structure.  That is the familiar condition for preserving $\cN=2$ supersymmetry~\cite{Becker:1996gj}.

\end{enumerate}

\subsection{Structures and extended supersymmetry}
As we have seen, for particular choices of flux we obtain vacua with various amounts of extended supersymmetry.  In this section we will make a more systematic study of the constraints that lead to $\cN = 2,3,4$, and we will also explore the massless spectrum of these theories.  To start, we note that every $\Sp(2)$ structure on $X\times\Xt$ compatible with the product metric takes the form
\begin{align}
\label{eq:SP2}
\cJ_A & = R_{Aa} j_a + \Rt_{A \dot a} \jt_{\dot a}~,
\end{align}
where $R$ and $\Rt$ are $3\times 3$ $\SO(3)$ matrices; it is an easy matter to check that these satisfy the defining relations of $\Sp(2)$ structure (see footnote~\ref{foot:sp2}).  To show that every $\Sp(2)$ structure takes this form, we just note that by raising an index on the $\cJ_A$ with the metric we should obtain the triplet of complex structures satisfying the familiar quaternionic algebra; that fixes the $\cJ_A$ in the form shown.

Similarly, the most general $\SU(4)$ structure on $X\times\Xt$ takes the form
\begin{align}
J & = y_A \cJ_A~,&
\Omega & = \ff{1}{2} (u_A \cJ_A)^2~.
\end{align}
Here $y_A$ is a real vector and $u_A$ is a complex vector such that the $3\times 3$ matrix
\begin{align*}
\begin{pmatrix} \Rea(u_1) & \Img (u_1) & y_1 \\ \Rea(u_2) & \Img (u_2) & y_2 \\ \Rea(u_3) & \Img (u_3) & y_3 \end{pmatrix} \in \SO(3)~.
\end{align*}
In particular, $y^T u = 0$, $2y^T y = u^\dag u = 2$, and $u^T u = 0$.  Setting $x = y^T R$, $t = u^T R$, and similarly for $\xt$ and $\ttld$, and
using~(\ref{eq:SP2}), we
can write the most general $\SU(4)$ structure on $X\times\Xt$ as
\begin{align}
J & = x_a j_a +\xt_{\dot a} \jt_{\dot a}~, &
\Omega & = t_a \ttld_{\dot a} j_a \jt_{\dot a}~,
\end{align}
where $x$ and $t$, as well as $\xt$ and $\ttld$ satisfy the same conditions as $y,u$.

\subsubsection*{$\cN\ge 2 $ supersymmetry}
Having taken care of the preliminaries, the analysis of the supersymmetry conditions is now straightforward.  To preserve at least $\cN =2$
we know that $G$ must be a primitive (2,2) form~\cite{Becker:1996gj} with respect to some $\SU(4)$ structure.  In other words,
\begin{align}
J \wedge G & = 0~, &\Omega \wedge G & = 0~,
\end{align}
and $G$ has no (1,3) or (3,1) components.  Applying the first two conditions to the general flux in~(\ref{eq:PTflux}), we obtain
\begin{align}
\label{eq:SU41}
J \wedge G & = 0~ && \iff & M \xt + (2A-C) x & =0~ &\text{and} ~&& x^T M + (2A-C) \xt^T & = 0~, \nonumber \\
\Omega \wedge G & = 0~ && \iff & t^T M \ttld & = 0~.
\end{align} 
To ensure no (1,3) or (3,1) components in $G$ we note that the harmonic $(3,1)$ forms on $X\times\Xt$ with respect to the chosen $\SU(4)$ structure are all
linear combinations of
\begin{align}
t_a j_a \wedge \xt_{\dot a} \jt_{\dot a}~, &&
t_a j_a \wedge\omega_{\dot \alpha}~, &&
x_a j_a \wedge\ttld_{\dot a} \jt_{\dot a}~, &&
\omega_\alpha  \wedge  \ttld_{\dot a} \jt_{\dot a}~.
\end{align}
So, our final requirement is that $G$ is annihilated by each of these terms.  This leads to the conditions
\begin{align}
\label{eq:SU42}
t^T M \xt & = 0~, &
x^T M \ttld & = 0~.
\end{align}
The vectors $x, \Rea(t), \Img(t)$ form an orthonormal basis, as do $\xt, \Rea (\ttld), \Img(\ttld)$, and by taking real and imaginary parts of~(\ref{eq:SU41}, \ref{eq:SU42}),
we find that in order for $G$ to be compatible with some $\SU(4)$ structure the matrix $M$ must be expressible as
\begin{align}
M & = S^T \begin{pmatrix} C-2A & 0 & 0 \\ 0 & \alpha & \beta \\ 0 & -\beta & \alpha \end{pmatrix} \St~,
\end{align}
where $S$ and $\St$ are $\SO(3)$ matrices.  This implies
\begin{align}
M M^T & = S^T \begin{pmatrix} (C - 2A)^2 & 0 & 0 \\ 0 & \alpha^2+\beta^2  & 0 \\ 0 & 0 & \alpha^2+\beta^2 \end{pmatrix} S~,
\end{align}
so that  $MM^T$ has $(C-2A)^2$ as an eigenvalue, and $MM^T$ has at least two equal eigenvalues.
These requirements can be easily translated into polynomial conditions on the parameters $A,B_{1},B_{2}, C, D_{1}, D_{2}$ that appear in~(\ref{eq:PTflux})~.

\subsubsection*{$\cN\ge 3$ supersymmetry}
The flux will be compatible with an $\Sp(2)$ structure if and only if $\cJ_{A} \wedge G = 0$ for $A=1,2,3$.  Using the $\cJ_{A}$ in~(\ref{eq:SP2}), this leads to
\begin{align}
M & = (C-2A) R^T \Rt~,
\end{align}
so that $MM^T = (C-2A)^2 \iden_{3}$.   Finally, to be compatible with $\Sp(1)\times\Sp(1)$ structure and therefore $\cN = 4$ supersymmetry, 
the condition on $M$ has to be true for all $R,\Rt$ in $\SO(3)\times\SO(3)$.  This forces $M=0$.

Note that the volume-threading term in the $G$-flux is proportional to $(C-2A)$.  This means that every $\cN \ge 3$ vacuum without a volume-threading
term necessarily has $M = 0$, so that it is actually preserving $\cN = 4$.

It is not obvious that we can choose an integral $G$-flux that both takes the $\cN=3$ form and satisfies the Bianchi identity without M2-branes.  Appendix~\ref{app:n3flux} shows this to be the case.

\subsection{Massless spectrum}
Like the existence of the vacuum, the massless spectrum also correlates nicely with the structure preserved by the flux.  We will not go into a detailed study
of the interactions and, for example, explicit expressions for the moduli space metric; this has been carried out at the supergravity level for the most general
flux compatible with minimal supersymmetry in~\cite{Prins:2015nda}.  Instead, we will just point out how the counting of massless degrees of freedom correlates
with the structure. 

\subsubsection*{Metric moduli}
The $58$--dimensional space of first-order deformations of an Einstein metric on $X$ can be parametrized in terms of a scalar parameter $x$
that corresponds to rescaling the total volume, as well as a $3\times 19$ matrix $\cX_{a\alpha}$:
\begin{align}
\label{eq:metricdeformation}
\delta j_a &= x j_a + \cX_{a\alpha} \omega_{\alpha}~,&
\delta \omega_\alpha &= x \omega_\alpha - j_a \cX_{a\alpha}~,&
\delta v &= 2xv~.
\end{align}
It is easy to see that this preserves the defining conditions:
\begin{align*}
\delta \left( j_a j_b - 2\delta_{ab} v E\right) & = 0~,&
\delta \left( j_a \omega_\alpha\right) & = 0~,&
\delta \left(\omega_\alpha \omega_\beta + 2 \delta_{\alpha\beta} v E\right) & = 0~.
\end{align*}
We have analogous expressions for the other K3 $\Xt$, except for tildes and dots.

Not all of these geometric deformation parameters correspond to three--dimensional massless modes:  a necessary condition is that the integral (and therefore rigid) flux $G$ satisfies the same conditions with respect to the deformed and undeformed $j_a$ and $\jt_{\dot a}$.  Since we parametrized $G$ in terms of the basis of self-dual
and anti-self-dual forms on $X$ and $\Xt$, this amounts to finding $\delta M$ and $\delta f$ in~(\ref{eq:PTflux}) such that under~(\ref{eq:metricdeformation}) $\delta G = 0$.
Plugging all of the variations into $G$ and demanding $\delta G = 0$, we obtain the following conditions:
\begin{align}
\delta M & = -(x+\xt) M~, &
\delta f & = -(x+\xt) f~,
\end{align}
and 
\begin{align}
(2A-C)(x-\xt) & = 0~, &
\cX^T M & = f \cXt^T~, &
M \cXt & = \cX f~.
\end{align}
The first two equations merely determine $\delta M$ and $\delta f$ and do not lead to interesting constraints.  On the other hand, the remaining three are
interesting.  First, we see that if $2A \neq C$ then $x = \xt$, so that while the overall volume modulus of $X\times \Xt$ remains massless, it is not possible to
tune the volumes of $X$ and $\Xt$ separately.  Thus, $2A=C$ is a necessary condition to be able to lift the vacuum to $7$ dimensions.

The remaining conditions are covariant with respect to the obvious $\GO(3)\times\GO(3)\times\GO(19)\times\GO(19)$ action on the $j_a,\jt_{\dot a}$ and $\omega_\alpha,\omegat_{\dot\alpha}$.
This means we can use singular value decomposition to bring $M$ and $f$ to canonical form:
\begin{align}
M &= \diag (\mu_1,\mu_2,\mu_3)~, &
f   & = \diag (\phi_1,\phi_2,\ldots, \phi_{19})~,
\end{align}
where the $\mu_a$ and $\phi_a$ are all non-negative (they are positive square roots of the eigenvalues of, respectively, $MM^T$ and $f f^T$.).  In this form the 
conditions on $\cX$ and $\cXt$ are written as
\begin{align}
\mu_a \cX_{a\alpha} & = \cXt_{a\alpha} \phi_\alpha~, &
\mu_a \cXt_{a\alpha} & = \cX_{a\alpha} \phi_\alpha~, &  \text{no sum on $a$ or $\alpha$.}
\end{align}
Generically these require $\cX = \cXt = 0$, but if some of the eigenvalues of $M$ match those of $f$, there are more solutions.  The number of
independent parameters is given by
\begin{align}
n(\cX,\cXt) &= 2 \dim \ker M \dim \ker f + \sum_{a,\alpha~| \mu_a \neq 0} \delta(\mu_a -\phi_a)~, \nonumber\\
  & = \dim \ker M \dim \ker f + \sum_{a,\alpha} \delta(\mu_a -\phi_a)~,
\end{align}
where $\delta(\mu_a -\phi_\alpha) = 1$ if $\mu_a = \phi_\alpha$ and is zero otherwise.
To understand this, note that the second line follows trivially from the first.  The first line merely says that if $\mu_a = 0$ then the equations require that
$\cX_a$ and $\cXt_{\dot a}$ both belong to $\ker f$; if for a fixed $a$ $\mu_a\neq 0$, then $\cXt_{a\alpha}$ is determined by $\cX_{a\alpha}$, and the latter satisfies
$(\mu_a - \phi_\alpha) \cX_{a\alpha}  = 0$.  

Including the constraints on the $x,\xt$, we find that the massless metric moduli are counted by
\begin{align}
\#(\text{metric moduli}) = \begin{cases}   2 + n(\cX,\cXt)~,  & 2A = C~, \\ 1+n(\cX,\cXt)~, & 2A \neq C~. \end{cases}
\end{align}

\subsubsection*{Massless vectors}
Fluctuations of $C$ give rise to massless vectors in three dimensions: 
$C = \cdots + \bV^I \Omega_I$, where the $\bV^I$ are three-dimensional vectors with field strengths $\bF^I = d\bV^I$, and the $\Omega_I$
are harmonic forms on $X\times \Xt$.  Inserting this into the M-theory action leads to a Chern-Simons mass term for the $\bV^I$ proportional to 
\begin{align}
\Delta\cL_3 = \bV^I \bF^J \int_{M_8} G \Omega_I \Omega_J~.
\end{align}
To explore the kernel of this mass term we write out
\begin{align}
V^I \Omega_I & = V^a_+ j_a + V^\alpha_- \omega_\alpha +\Vt^{\dot a}_+ \jt_{\dot a} + \Vt^{\dot\alpha}_- \omegat_{\dot\alpha}~
\end{align}
and combine these components into a $44$--dimensional vector
\begin{align}
\mathbb{V}^T = (V_+^T \,\,  \Vt_+^T \,\, V_-^T \,\,  \Vt_-^T)~,
\end{align}
and similarly for the field-strengths, which are packaged in a vector $\mathbb{F}$.  With a little bit of algebra we find
\begin{align}
\Delta \cL_3 = 4 v\vt \mathbb{V}^T \mathbb{M} \mathbb{F}~,
\end{align}
where
\begin{align}
\mathbb{M} = \begin{pmatrix} \mathbb{M}_+ & 0 \\ 0 & \mathbb{M}_- \end{pmatrix}~,
\end{align}
and
\begin{align}
\mathbb{M}_+ & = \begin{pmatrix} (2A-C)\iden_3 & M \\ M^T& (2A-C) \iden_{3} \end{pmatrix}~,&
\mathbb{M}_-  & = \begin{pmatrix} (C-2A) \iden_{19} & f \\ f^T & (C-2A) \iden_{19} \end{pmatrix}~.
\end{align}
So, the number of massless vectors is $\dim \ker\mathbb{M}_+ + \dim \ker \mathbb{M}_-$.
The latter depend on the value of $(2A-C)$:
\begin{align}
\dim \ker\mathbb{M}_+ & = \begin{cases}    2 \dim\ker M & 2A=C \\ \dim\ker(M^TM -(2A-C)^2 \iden_3) & 2A\neq C \end{cases}~, & \nonumber\\
\dim \ker\mathbb{M}_- & = \begin{cases}    2 \dim\ker f & 2A=C \\ \dim\ker(f^Tf -(2A-C)^2 \iden_{19}) & 2A\neq C \end{cases}~, & \nonumber\\
\end{align}

\subsubsection*{Summary for $\cN=2,3,4$}
We now combine the previous results with the constraints on $M$ and $\cN$ found in the previous section.   In each case we will find
a result consistent with the three-dimensional multiplet structure for the particular $\cN$.
\begin{enumerate}
\item $\cN=4$.  This requires $M=0$ and therefore leads to
\begin{align}
\#(\text{metric moduli} ) & = 2 + 6 \dim\ker f~, \nonumber\\
\#(\text{massless vectors}) & = 6+2\dim\ker f~.
\end{align}
Recall that the massless vector and hyper multiplets of $\cN=4$ each contain $4$ scalar degrees of freedom; this is consistent with the total
number of massless scalars obtained here (which is in fact divisible by $8$).  We do not expect quantum corrections to lift any of these massless degrees of 
freedom.
\item $\cN=3$.    In this case $\mu_a = |2A-C| \neq 0$ for $a=1,2,3$, and therefore
\begin{align}
\label{eq:N3count}
\#(\text{metric moduli} ) & =  1+3 \dim \ker\{f^T f -(2A-C)^2\iden_{19}\}~, \nonumber\\
\#(\text{massless vectors}) & = 3 + \dim \ker\{f^T f - (2A-C)^2 \iden_{19}\}~.
\end{align}
Since the massless supermultiplets for $\cN =3 $ have exactly the same structure as the more familiar $\cN=4$ multiplets~\cite{deWit:1992psp}, we expect
that the total number of scalars is divisible by $4$, and indeed it is.  The moduli space of $\cN=3$ theories is quaternionic~\cite{deWit:1992psp}, and we suspect but
have not checked that, as in the $\cN=4$ case, supersymmetry is sufficient to rule out quantum corrections that might lift these degrees of freedom.

\item $\cN=2$.  In this case we expect quantum corrections to lift some of the classically massless fields, so our results are merely upper bounds on
the massless spectrum.  The content of $\cN=2$ chiral and vector multiplets easily follows by reduction from $d=4$ $\cN=1$ multiplets, and each massless multiplet contains
two scalar degrees of freedom.  Based on the analysis above, we find the following massless spectrum; in each case we do find the expected even number of scalars.
\begin{enumerate}
\item  The generic case is when $\mu_1 = |2A-C|\neq 0$ and $0 < \mu_2 = \mu_3 \neq \mu_1$.
\begin{align}
\#(\text{metric moduli} ) & =  1+\dim\ker\{f^T f - \mu^2_1\iden\} + 2 \dim\ker\{f^Tf-\mu^2_2 \iden\}~, \nonumber\\
\#(\text{massless vectors}) & = 1 + \dim \ker\{f^T f - \mu^2_1 \iden\}~.
\end{align}
\item  A less generic possibility $\mu_1 = |2A-C| \neq 0$ and $\mu_2 = \mu_3 =0$ leads to
\begin{align}
\#(\text{metric moduli} ) & =  1+4 \dim \ker f + \dim\ker\{f^T f- \mu^2_1 \iden\}~, \nonumber\\
\#(\text{massless vectors}) & = 1 + \dim \ker\{f^T f - \mu^2_1 \iden\}~.
\end{align}
\item The final possibility, $\mu_1 = |2A-C| =0$ and $0 < \mu_2 =\mu_3$, leads to
\begin{align}
\#(\text{metric moduli} ) & =  2+2 \dim \ker f + 2\dim\ker\{f^T f- \mu^2_2 \iden\}~, \nonumber\\
\#(\text{massless vectors}) & = 2 + 2\dim \ker f~.
\end{align}
\end{enumerate}

\end{enumerate}

\section{Heterotic 3d compactifications} \label{s:heterotic}
The preceding sections identified and studied a large class of M-theory vacua based on the relatively simple geometry of K3$\times$K3.  In this section we will consider potential dual heterotic descriptions of these vacua in three dimensions.  There are many examples of dual pairs based on the $7$--dimensional duality between a heterotic string on $T^3$ and M-theory on K3.  For instance, we expect to be able to find M-theory descriptions of heterotic backgrounds satisfying the following two conditions:
\begin{enumerate}
\item the three--dimensional gauge group is abelian;
\item the compactification manifold $X_7$ is a principal $T^3$ fibration over K3, with the bundle obtained by a combination of Wilson lines and a pull-back of a bundle from the base K3 geometry.
\end{enumerate}
These geometries have a lift to $7$ dimensions, and fiberwise duality with M-theory on K3 should make sense.  

On the other hand, as we already saw, M-theory solutions with $G$-flux that threads the volumes of the K3s do not have simple heterotic duals.  We outlined some of the challenges of finding the duality in terms of the massive $7$--dimensional theory in section~\ref{s:M7}.  We will now consider the problem directly in $3$ dimensions, and we will show that there are no heterotic geometries that lead to exactly $\cN = 3$ supersymmetry in three dimensions:  a solution with $6$ supercharges actually preserves $8$ or $16$ supercharges.  

\subsection{Review of heterotic $\GG_2$ geometry}
Consider a three--dimensional compactification of the heterotic string with $\cN \ge 1$ on a seven--dimensional compact manifold $X_7$.  In order to discuss spinors and their properties on $X_7$ let us first fix a basis for the Clifford algebra.\footnote{A thorough and readable review of $\GG_2$--structure compactification is given in~\cite{Becker:2014rea}; we follow it in a number of conventions, including that for the spinors.}  

\subsubsection*{Clifford algebra on $X_7$}
We choose the $\Gamma_i$, $i=1,\ldots, 7$ to be a pure imaginary antisymmetric basis satisfying 
\begin{align*}
\AC{\Gamma_i}{\Gamma_j} = 2g_{ij}~.
\end{align*}  
The matrices $\{\iden,  i\Gamma_{ijk}\}$ are real symmetric, while $\{i\Gamma_i, \Gamma_{jk}\}$ are real anti-symmetric.  Together they span the Clifford algebra:  given a non-zero real spinor $\ep_0$ a basis of spinors is $\{\ep_0, i \Gamma_i \ep_0\}$.  That is, we have the completeness relation
\begin{align}
\label{eq:completeness}
\Gamma^i \ep_0 \ep_0^t \Gamma_i + \ep_0 \ep^t_0 = \iden_{8}~.
\end{align}
In the usual way we define $\Gamma^{i_1 \cdots i_k} = \frac{1}{k!} \Gamma^{[i_1} \Gamma^{i_2} \cdots \Gamma^{i_k]}$, and we lower and raise the (co)tangent space indices with the metric $g_{ij}$ and its inverse $g^{ij}$.


\subsubsection*{Minimal supersymmetry requirements}

Minimal supersymmetry requires that the geometry satisfies the following conditions.\footnote{The general result goes back to~\cite{Strominger:1986uh}; applications to $X_7$ may be found in, for instance, \cite{Friedrich:2001yp,Gauntlett:2003cy}.}
\begin{enumerate}
\item The gauge bundle $\cP \to X_7$ has structure group in $\Spin(32)/\Z_2$ or $\GE_8\times\GE_8$ and satisfies the heterotic Bianchi identity in integral cohomology.
\item The vanishing of the gravitino variation requires that $X_7$ admits a $\nabla^-$--constant spinor $\ep_0$. The $\nabla^-$ connection is  the Levi-Civita connection twisted by the $3$--form $H$: 
\begin{align*}
\left(\Gamma^{-}\right)^{l}_{jk} = g^{li}\left( \ff{1}{2} \left[ g_{ji,k} + g_{ki,j} - g_{jk,i}\right] -\ff{1}{2} H_{ijk}\right) = \Gamma^{l}_{jk} - \ff{1}{2} H^{l}_{~jk}~.
\end{align*}
This means $X_7$ has $\GG_2$ structure.

\item  The vanishing of the dilatino variation requires
\begin{align}
\label{eq:dilatino}
\left[\p_i \vphi \Gamma^i -\ff{1}{12} H_{ijk} \Gamma^{ijk}\right] \ep_0 = 0~.
\end{align}
Here $\vphi$ is the dilaton field.
\item The gauge curvature $\cF$ annihilates the spinor:   $\cF_{ij} \Gamma^{ij} \ep_0 = 0$.
\item The Bianchi identity has a solution in the formal $\alpha'$ expansion~\cite{Melnikov:2014ywa}.  
\end{enumerate}
Conditions 2,3, and 4 will be sufficient for our purposes, but any putative solution must satisfy all of these necessary conditions.

The existence of a $\nabla^-$--constant spinor $\ep_0$ implies the existence of $\nabla^-$--constant associative and co-associative forms
\begin{align}
\Phi_{ijk} &= i \ep_0^T \Gamma_{ijk} \ep_0~,&
\Psi_{ijkl} & = \ep_0^T \Gamma_{ijkl} \ep_0~. 
\end{align}
The metric $g$ relates these two by $\ast_g \Phi = \Psi$, and $\Phi\wedge \ast_g \Phi = 7 \dVol_g(X_7)$.  Moreover, we have the 
helpful relations
\begin{align}
\label{eq:helpful}
\Gamma_{ij} \ep_0 & = -i \Phi_{ijk} \Gamma^k \ep_0~,&
i\Gamma_{ijk} \ep_0 & = \Phi_{ijk} \ep_0 - i \Psi_{ijkl} \Gamma^l \ep_0~.
\end{align}
The $\Phi$ and $\Psi$ obey a number of useful relations summarized in appendix A of~\cite{Becker:2014rea}.  We will find use for two of these:
\begin{align}
\label{eq:PsiPhi}
\Psi_{ijnm} \Phi^{klm} &= 6 \delta^{[k}_{[i} \Phi_{jn]}^{~~~l]}~, &
\Phi_{ijk} \Phi_{lm}^{~~~k}  &= g_{il} g_{mj} - g_{im} g_{lj} - \Psi_{ijlm}~.
\end{align}

Turning the construction around, suppose $X$ has a $\GG_2$ structure, i.e. a non-degenerate $3$-form $\Phi$ that in a local orthonormal frame $\{e^i\}_{i=1,\ldots,7}$ with respect to metric $g$ has the canonical form
\begin{align}
\label{eq:algebraicg2}
\Phi &=   e^{246} -e^{235} -e^{145} - e^{136} + e^{127} + e^{347} + e^{567}~,\nonumber\\
\ast \Phi & = e^{1234} + e^{1256} +e^{3456} + e^{1357} -e^{1467} -e^{2367} - e^{2457}~.
\end{align}
We use a condensed notation, where we omit the wedge symbol when it is unlikely to cause confusion, and we collapse labels on products of $1$-forms; thus, $e^{246} = e^2 \wedge e^4 \wedge e^6$, etc.

The necessary and sufficient conditions to satisfy conditions 2 and 3 are that, in addition to the
algebraic conditions of~(\ref{eq:algebraicg2}), we also have the differential conditions
\begin{align}
\label{eq:differentialg2}
\Phi \wedge d \Phi & = 0~,&
d\left[ e^{-2\vphi} \ast \Phi\right] & = 0~,&
\ast H &= e^{2\vphi} d\left[ e^{-2\vphi} \Phi\right]~.
\end{align}
Note that the last one determines the torsion $H$, and the last two involve the dilaton $\vphi$.

\subsection{Extended supersymmetry :  conditions on $X_7$} \label{ss:extended}
In order to have extended supersymmetry in $d=3$, $X_7$ must admit additional linearly independent $\nabla^-$--constant spinors.  Suppose there are $p+1$ such linearly independent spinors $\{\ep_0, \ep_1, \ldots,\ep_p\}$.  Let $A= 1,\ldots,p$ index the ``extra'' spinors.   Since $\{i \Gamma_i \ep_0, \ep_0\}$ are a complete basis, we can find vector fields $V_A^i$ and functions $u^A$ so that
\begin{align}
\ep_A = i V_A^i \Gamma_i \ep_0 + u^A \ep_0~
\end{align}
for each $A$.   Covariant constancy of $\ep_0$ requires $\nabla^- V_A = 0$ and $\nabla^- u_A = 0$; the latter means that the $u_A$ are just constants; we can  set $u^A=0$ without loss of generality \cite{Kaste:2003zd}.

We conclude that extended supersymmetry requires $X_7$ to admit  of $\nabla^-$--constant vector fields.  Conversely, given $p$ linearly-independent $\nabla^-$--constant vector fields
$V_A$, we can construct $p$ additional spinors $\ep_A$.   We can take the $V_A$ to be orthonormal.\footnote{It is not hard to show that the $V_A$ are Killing vectors; moreover their commutator is determined by a pairwise contraction with the torsion $H$. }

 The reader may recall that any compact $\GG_2$ structure manifold  admits $3$ nowhere vanishing vector fields which reduce the structure further to $\SU(2)$ \cite{MR1484591}.  However, we stress that the supersymmetry conditions are stronger: the vectors must be annihilated by $\nabla^-$.

\subsubsection*{Constraints on the number of vectors}
Suppose that $X_7$ satisfies the minimal supersymmetry conditions and admits exactly $p$ linearly independent $\nabla^-$--constant vectors $V_A$.
We will now show that the number of vectors $V_A$, $A=1,\ldots, p$ can only take on specific values:  $p \in \{0,1,3,7\}$.  Realizations of each of these cases are well known.
\begin{enumerate}
 \item $p=0$ corresponds to an irreducible $X_7$ --- this is minimally supersymmetric and exemplified by, for example, one of Joyce's manifolds of $\GG_2$ holonomy~\cite{Joyce:2000cm} (standard embedding for the gauge bundle is a standard solution of the other supersymmetry constraints).
 \item $p=1$, which leads to $\cN = 2$ in three dimensions, is also familiar:  for instance we can take $X_7 = X_6 \times S^1$, where $X_6$ is a Calabi-Yau $3$--fold; more generally, we can take $X_7$ to be a principal circle bundle over $X_6$.
 \item $p=3$, which leads to $\cN = 4$ in three dimensions can be obtained from $X_7 = \text{K3} \times T^3$; again, it is easy to make more general solutions by fibering the $T^3$ over K3.
 \item $p=7$, which leads to $\cN = 8$ in three dimensions can be obtained by taking $X_7 = T^7$.
\end{enumerate}

\subsubsection*{Two vectors imply a third}
Suppose we have two vectors $V_A$, $A=1,2$.  Given these, we can construct the dual $1$-forms $\Theta^A$, and we can also find a third $1$-form
\begin{align}
\Theta^3 &= V_1 \llcorner V_2 \llcorner \Phi~.
\end{align}
The $\llcorner$ denotes contraction of the vector field into the form:  given a $k$--form $\omega = \frac{1}{k!} \omega_{i_1 \cdots i_k} dx^{i_1} \cdots dx^{i_k}$, the $k-1$--form $V\llcorner \omega$ is
\begin{align*}
V\llcorner \omega = \frac{1}{(k-1)!} V^{i_1} \omega_{i_1 i_2 \cdots i_{k}} dx^{i_2} \cdots dx^{i_k}~.
\end{align*}
By construction $\Theta^3$ is $\nabla^-$--constant and annihilated by $V_1$ and $V_2$.  Hence, if $\Theta^3 \neq 0$, its dual $V_3$ will be a third $\nabla^-$--constant vector linearly independent from $V_1$ and $V_2$.

To show that $\Theta^3$ is non-zero, we compute its norm:
\begin{align}
\| \Theta^3 \|^2 = V_1^i V_2^j \Phi_{ijk}  V_1^l V_2^m \Phi_{lmn} g^{km} = V_1^i V_1^l V_2^j V_2^m \Phi_{ijk} \Phi_{lm}^{~~~k}~.
\end{align}
Using~(\ref{eq:PsiPhi}) we find
(recall that the $V_A$ are orthonormal by assumption)
\begin{align}
\|\Theta^3 \|^2 = \|V_1\|^2 \|V_2\|^2 = 1~.
\end{align}
Thus, if $X_7$ has $p\ge 2$  $\nabla^-$--constant vectors, then $p\ge 3$.

\subsubsection*{From $4$ to $7$ vectors}
We will now show that if $p\ge 4$, then $p = 7$.  Suppose that we have exactly $p$  $\nabla^-$--constant vectors $V^A$ and their dual $1$--forms $\Theta_A$.
We can choose all of these to be orthonormal and in any patch complete the basis with some $1$-forms $e^\alpha$, $\alpha = 1, \ldots, 7-p$.  The Hodge star then decomposes
as $\ast_7 = \ast_p \ast_{7-p}$, and the $3$-form $\Phi$ is
\begin{align}
\Phi = \Phi^{(3)} + \Phi^{(2)}_{A} \Theta^A + \ff{1}{2} \Phi^{(1)}_{AB} \Theta^{AB} + \ff{1}{3!} \Phi^{(0)}_{ABC} \Theta^{ABC}~,
\end{align}
where the $\Phi^{(s)}$ are $s$--forms constructed from the $e^\alpha$:
\begin{align}
\Phi^{(0)}_{ABC}& = \Theta^{ABC} \llcorner \Phi ~, \nonumber\\
\Phi^{(1)}_{AB} & =  \Theta^{AB} \llcorner \Phi- \ff{1}{2} \Phi^{(0)}_{ABC} \Theta^C~, \nonumber\\
\Phi^{(2)}_{A} & =  \Theta^{A} \llcorner \Phi- \Phi^{(1)}_{AB} \Theta^B - \ff{1}{2} \Phi^{(0)}_{ABC}\Theta^{BC}~,
\end{align} 
and $\Phi^{(3)}$ is found by taking the difference of these terms with $\Phi$.  Clearly the $\Phi^{(s)}$ are $\nabla^-$--constant.
In particular, the dual $\Phi^{(1)}$, if non-zero, would yield an additional vector that is linearly independent from the $\Theta^A$.  So,
we set $\Phi^{(1)} = 0$ and work with
\begin{align}
\Phi &= \Phi^{(3)} + \Phi^{(2)}_{A} \Theta^A  + \ff{1}{3!} \Phi^{(0)}_{ABC} \Theta^{ABC}~, \nonumber\\
\ast \Phi  & = (\ast_{7-p} \Phi^{(3)}) (\ast_p 1)
+ (\ast_{7-p} \Phi^{(2)}_A) (\ast_p \Theta^A)
+\ff{1}{3!}(\ast_{7-p} \Phi^{(0)}_{ABC}) (\ast_p \Theta^{ABC}) \nonumber\\[2mm]
\Psi & = \Psi^{(4-p)} (\ast_p 1) + \Psi^{(5-p)}_A (\ast_p \Theta^A) + \ff{1}{3!} \Psi^{(7-p)}_{ABC} (\ast_p \Theta^{ABC})~.
\end{align}
Since $\Psi = \ast \Phi$, the last line is merely convenient notation for the contents of the second one.  By the same arguments as above,
the $\Psi^{(7-p-s)}$ are $\nabla^-$--constant and linearly independent from the $\Theta^A$.

Now consider the possibility $p=4$.  This requires $\Phi^{(2)}_A =0$, since otherwise $\Psi^{(1)}_A$ yields an additional
$1$--form.  On the other hand, since $\wedge^3 \R^4 = \R$ we can write $\Phi^{(0)}_{ABC}  = \ep_{ABCD} Y^D$ for some constants $Y^D$,
but this contradicts non-degeneracy of $\Phi$ because  $\Phi$ is annihilated by $\sum_{A} Y^A V_A$.

Similarly, $p=5$ is not compatible with a non-degenerate metric.  To see this, recall that $\Phi$ determines the metric as
\begin{align}
g_{ij} = \frac{1}{144} \ep^{klmnpqr} \Phi_{ikl} \Phi_{jmn}\Phi_{pqr}~.
\end{align}
If $p=5$, then $\Phi$ is given by
\begin{align}
\Phi = e^{12} k_A \Theta^A + \Phi^{(0)}_{ABC} \Theta^{ABC}~,
\end{align}
and a moment's thought shows the contradiction: on one hand we assumed without loss of generality that $\{e^1, e^2, \Theta^1,\ldots, \Theta^5\}$ is an
orthonormal basis, but on the other hand from the explicit formula for $g$ we have $g_{11} = 0$.

Finally, $p\ge 6$ implies $p=7$ because otherwise $\Phi$ is  annihilated by the dual of $e^1$. 

\subsection{A no-go theorem}
As we just argued, the geometry of $X_7$ admits exactly $p$ $\nabla^-$--constant vectors $V_A$ only if $p \in \{0,1,3,7\}$. Naively such $X_7$ lead to $\cN = \{1,2,4,8\}$ supersymmetry in three dimensions.  Of course this requires that the remaining supersymmetry conditions are obeyed with $\ep_0$ replaced by the corresponding $\ep_A$, and it may be that this only holds for some $k<p$ spinors.  This would lead to extended supersymmetry with $\cN  = k+1$.  We will now prove the following no-go result:  if $p>1$ then $k >2$, so that a solution with  $\cN \ge 3$ necessarily has $\cN \ge 4$.  Similarly, $\cN \ge 5$ implies $\cN = 8$.

To get started, we note that~(\ref{eq:dilatino}) holds if and only if
\begin{align}
\label{eq:formdilatino}
\Phi \llcorner H & = 0~,&
2d\vphi & = - H \llcorner\Psi~.
\end{align}
To show this we apply the completeness relation~(\ref{eq:completeness}) to ~(\ref{eq:dilatino}), which shows the latter to
be equivalent to
\begin{align*}
0 &= \ep_0^T  \left[ \nabla^i \vphi \Gamma_i - \ff{1}{12} H^{ijk} \Gamma_{ijk} \right] \ep_0~,&
0 &= \ep_0^T \Gamma_m \left[ \nabla^i \vphi \Gamma_i - \ff{1}{12} H^{ijk} \Gamma_{ijk} \right] \ep_0 ~.
\end{align*}
Since $\Gamma_i$ is antisymmetric, the first equation is the statement $\Phi \llcorner H = 0$; using~(\ref{eq:helpful})
the second condition leads to $2d \vphi  = -H\llcorner\Psi$.  

Similarly, applying~(\ref{eq:helpful}) to the gaugino variation, we learn
that 
\begin{align}
\label{eq:formgaugino}
\cF^{ij} \Gamma_{ij} \ep_0 &= 0  & \iff  & &
\cF \llcorner \Phi &  = 0~ & \iff & &
\cF & = \cF \llcorner\Psi~.
\end{align}
The third relation follows from the second by contracting $\cF \llcorner \Phi$ into (the non-degenerate) $\Psi$ and using~(\ref{eq:PsiPhi})~.
With these preparations in hand, we assume minimal supersymmetry, and we turn to extended supersymmetry.

\subsubsection*{The gravitino variation for $\ep_A$}
The existence of the spinors $\ep_A$ yields extra $\GG_2$ structures:
\begin{align}
\Phi^A_{ijk} &= i \ep^T_A \Gamma_{ijk} \ep_A~, &
\Psi^A_{ijkl} &= \ep^T_A \Gamma_{ijkl} \ep_A~.
\end{align}
Using $\ep_A = i V_A^n \Gamma_n\ep_0$ we can also write this as
\begin{align}
\Phi^A_{ijk} =  i V^m_A V^n_A \ep_0^T \Gamma_m \Gamma_{i}\Gamma_{j}\Gamma_{k} \Gamma_n \ep_0~.
\end{align}
By commuting the $\Gamma_m$ through $\Gamma_{i}\Gamma_j\Gamma_k$, we obtain an elegant form for $\Phi^A$:
\begin{align}
\Phi^A = 2 \Theta^A \wedge (V_A \llcorner\Phi) - \Phi~.
\end{align}
In other words, to obtain $\Phi^A$ from $\Phi$ we write out $\Phi$ in a $\Theta$ expansion, and we flip the sign of every
term that does contain $\Theta^A$.  The $\Phi^A$ will be $\nabla^-$--constant since $\Phi$ and $\Theta^A$ are $\nabla^-$--constant.  Note that
\begin{align}
\nabla^- \Theta^A &= 0~ & \implies &&  d\Theta^A &= V^A \llcorner H~.
\end{align}

\subsubsection*{The dilatino variation for $\ep_A$}
The dilatino variation will vanish for $\ep_A$ provided that
\begin{align}
V_A^m \left(\nabla^i \vphi \Gamma_i -\frac{1}{12} H^{ijk} \Gamma_{ijk} \right) \Gamma_m \ep_0 & = 0~.
\end{align}
Since it vanishes for $\ep_0$, we can replace this with the anti-commutator 
\begin{align}
V_A^m \Bigl\{ \left(\nabla^i \vphi \Gamma_i -\frac{1}{12} H^{ijk} \Gamma_{ijk} \right), \Gamma_m\Bigr\} \ep_0 & = 0~,
\end{align}
and some Clifford algebra manipulations, together with~(\ref{eq:helpful}), reduce this to
\begin{align}
V_A \llcorner d\vphi & = 0~, &
(V_A \llcorner H) \llcorner\Phi & = 0~.
\end{align}

\subsubsection*{The gaugino variation for $\ep_A$}
Finally, we have
\begin{align}
\cF^{ij}\Gamma_{ij} \ep_A = i V_A^m \cF^{ij} \CO{\Gamma_i\Gamma_j}{\Gamma_m} \ep_0 = 4i V_A^m \cF^{ij} g_{jm}\ep_0~.
\end{align}
So, the vanishing of the gaugino variation for $\ep_A$ reduces to
\begin{align}
V_A \llcorner \cF & = 0~.
\end{align}

\subsubsection*{$\cN\ge 3$ implies $\cN\ge 4$}
Suppose that all of the supersymmetry conditions are satisfied by $\ep_0$ and $\ep_A$ for $A=1,2$.  We will call these
the $\cN = 3$ supersymmetry conditions.  From the results above
we know that there exists a third $\nabla^-$--constant spinor
\begin{align}
\ep_3 & = i V_2^m \Gamma_m \ep_1 = i V_3^m\Gamma_m \ep_0~, &
V_3^m & = V_2^i V_1^j  \Phi_{ij}^{~~m}~.
\end{align}
We will now show that $\ep_3$ also yields a supersymmetry.  

Let us start with the gaugino variation.  Minimal supersymmetry requires $\cF = \cF \llcorner\Psi$, so
\begin{align}
V_3^m \cF_{mn} = \ff{1}{2} V_2^k V_1^l \Phi_{kl}^{~~m} \cF^{ij}\Psi_{ijmn} = -\ff{1}{2} \Theta^2_k \Theta^1_l \cF^{ij} \Psi_{ijnm} \Phi^{klm}~.
\end{align}
Using~(\ref{eq:PsiPhi}) we then obtain
\begin{align}
V_3^m \cF_{mn} = -\Theta_k^2 \Theta_l^1 \cF^{ij} 
\left(\delta^{[k}_{[i} \Phi_{jn]}^{~~~l]}+\delta^{[k}_{[j} \Phi_{ni]}^{~~~l]}+\delta^{[k}_{[n} \Phi_{ij]}^{~~~l]} \right) =0~.
\end{align}
The last equality follows because every term in the sum is proportional to either $V_1 \llcorner \cF$, $V_2 \llcorner \cF$, or to $\cF \llcorner \Phi$,
and all of these are zero by the $\cN=3$ conditions.

Next, we consider the term $V_3 \llcorner d\vphi$ that arises from the dilatino variation.  Using minimal supersymmetry we have
\begin{align}
-2 V_3 \llcorner d\vphi = V_3 \llcorner (H\llcorner \Psi)~.
\end{align}
In components we have
\begin{align}
-2 V_3\llcorner d\vphi & = \ff{1}{2} \Theta^2_p \Theta^1_q \ff{1}{3!} H^{ijk} \Psi_{ijkm} \Phi^{pqm}~,
\end{align}
and~(\ref{eq:PsiPhi}) allows us to rewrite this as
\begin{align}
-2 V_3\llcorner d\vphi & = \ff{1}{2} V_1 \llcorner \left[ (V_2 \llcorner H) \llcorner \Phi\right]
-\ff{1}{2} V_2 \llcorner \left[ (V_1 \llcorner H) \llcorner \Phi\right] = 0~.
\end{align}
The last equality follows because each square bracket is zero by $\cN = 3$ conditions.

Finally, we need to show that $(V_3\llcorner H) \llcorner \Phi = 0$.  This requires more details on the structure of $\Phi$ and $H$.\footnote{At the level of
representation theory the comparative difficulty can be traced to the fact that $H$ has components in both $\rep{27}$ and $\rep{7}$ of  $\wedge^3 T^\ast_X$
under the $\GG_2$ structure decomposition.}   
The first ingredient is the form of $\Phi$ with $p=3$ vectors. As we show in the appendix, we have
\begin{align}
\label{eq:PhiK3T3}
\Phi & = \omega_1 \Theta^1 +\omega_2\Theta^2 + \omega_3\Theta^3  + \Theta^{123}~,
\end{align}
where $\omega_A = \ff{1}{2} M_{Aij} e^i \wedge e^j$ are three self-dual $2$--forms that satisfy the $\SU(2)$ structure relations $\omega_A \wedge \omega_B = 2\delta_{AB}e^{1234}$.
This implies that $\Psi$ is given by
\begin{align}
\Psi = \ff{1}{2} \omega_A \ep_{ABC}\Theta^{BC} + e^{1234}~.
\end{align}
Next, we obtain constraints on $H$.  The $H$ flux has  a general expansion
\begin{align}
H = H^{(3)} + H_A^{(2)} \Theta^A + \ff{1}{2} H_{AB}^{(1)} \Theta^{AB} + H^{(0)} \Theta^{123}~,
\end{align}
and minimal supersymmetry requires
\begin{align}
\label{eq:HintoPhi}
0 & = H\llcorner\Phi = H_A^{(2)} \llcorner \omega_A + H^{(0)}~.
\end{align}
A short computation shows that the $\cN=3$ supersymmetry conditions imply $H_{AB}^{(1)} = 0$ for all $A$ and $B$, 
while 
\begin{align}
\label{eq:HA2}
H_A^{(2)} \llcorner \omega_B = - \delta_{AB} H^{(0)}~ \qquad A=1,2,\qquad\qquad  B=1,2,3~.
\end{align}
There are further constraints on $H$ from the minimal supersymmetry conditions.
First, since $H$ determines $d\Theta^A$ via
\begin{align}
d\Theta^A = V_A \llcorner H = H_A^{(2)} + \ff{1}{2} H^{(0)} \ep_{ADE} \Theta^{DE}
\end{align}
we see that
\begin{align}
d\Phi = \omega_A H_A^{(2)}  + \{\text{terms with at least one $\Theta$}\}~.
\end{align}
Therefore, $\Phi \wedge d\Phi = 0$ implies\footnote{The second condition follows from the first because $\omega_A = \ast \omega_A$. }
\begin{align}
H_A^{(2)} \omega_A &= 0 & & \iff & H_A^{(2)} \llcorner \omega_A = 0~.
\end{align}
 Combining this result with
~(\ref{eq:HintoPhi}), we conclude that $H^{(0)} = 0$, so that $V_A \llcorner H = H_A^{(2)}$.

For our last machination we note that since $H\llcorner\Psi = -2 d\vphi$, and $V^A \llcorner d\vphi = 0$, $H\llcorner \Psi$
cannot have any $\Theta$ components.  On the other hand, we have
\begin{align}
H \llcorner \Psi = H^{(3)} \llcorner e^{1234} - H_A^{(2)} \llcorner \omega_B \ep_{ABC}\Theta^C~.
\end{align}
The latter terms vanish if and only if 
\begin{align}
H_A^{(2)} \llcorner\omega_B  = H_B^{(2)} \llcorner\omega_A
\end{align}
for all $A,B$.  But, combining this with~(\ref{eq:HA2}) and $H^{(0)} = 0$, we finally have
\begin{align}
H_A^{(2)} \llcorner\omega_B = 0~
\end{align}
for all $A$ and $B$.  So, at last, $(V_A \llcorner H) \llcorner\Phi = 0$ for all $A$, and, as promised, $\cN\ge 3$ implies $\cN\ge 4$.

Incidentally, $H^{(0)} = 0$ also implies that all three vectors $V_A$ commute, so the $\cN = 4$ solutions are all of the form of
a $T^3$ bundle over a hyper-Hermitian surface.  Just as in the analogous case of $d=4$ $\cN=2$ compactifications~\cite{Melnikov:2012cv}, we
expect that the most general geometric solution of this form is indeed a $T^3$ bundle over a K3.

\subsubsection*{$\cN\ge 5$ implies $\cN = 8$}
Finally, we show that a compactification with $\cN \ge 5$ necessarily preserves maximal supersymmetry.  
By assumption of $\cN\ge 5$, we have $\ep_0$ and $\ep_A = iV_A^m\Gamma_m \ep_0$, with $A=1,2,3$,
as well as $\ep_4 = i V_4^m\Gamma_m \ep_0$ that solve the supersymmetry constraints.    We also know that $X_7$ admits three more independent $\nabla^-$--constant vectors $\Vt_{a}$, with $a=5,6,7$.  Without loss of generality we take the $\Vt_a$ orthonormal and orthogonal to the $V_A$; we define their dual forms $\Thetat^a$.

From above we know that for $\cN \ge 4$ $\Phi$ takes the form
\begin{align*}
\Phi =  \omega_A \Theta^A + \Theta^{123}~,
\end{align*}
where the $\omega_A$ are self-dual and satisfy $\omega_A \wedge \omega_B = 2 \delta_{AB} \Theta^4\Thetat^{567}$.  The conditions on $\omega_A$ imply that 
\begin{align}
\omega_A = U_{A a} \left[  \Theta^4 \Thetat^a + \ff{1}{2} \ep^{abc} \Thetat^b \Thetat^c\right]~,
\end{align}
with $U_{Aa} U_{Ba}  = \delta_{AB}$.  Hence $\Theta'^A = \Theta^4 \llcorner \omega_A$ are three orthonormal $\nabla^-$--constant $1$-forms that are also orthogonal to $\Theta^{1},\ldots,\Theta^{4}$.  The dual vectors $V'_A$ complete the $V_A$ to a basis for $T_X$.
Moreover, we have 
\begin{align}
\Theta'^A = V_A \llcorner (V_4 \llcorner \Phi)~,
\end{align}
and therefore the arguments we gave in the previous section guarantee that the spinors $\ep'_A$ constructed using the vectors $V'_A$ satisfy all of the supersymmetry conditions and generate three additional supersymmetries.


\acknowledgments IVM would like to thank the organizers of ``Mathematics of String Theory'' program for providing financial support and a very stimulating environment at the Institut Henri Poincar{\'e} for this work;  IVM and RM  also thank the Enrico Fermi Institute for hospitality.    We thank P.~S.~Aspinwall, D.~Morrison, D.~Prins, R.~Plesser and S.~Theisen for useful discussions. We are supported, in part, by the James Madison University's Faculty Assistance Grant (IVM), Agence Nationale de la Recherche under grant 12-BS05-003-01 (RM), and NSF Grant No. PHY-1316960 (SS).

\appendix

\section{ Three vectors on $X_7$ and constraints on $\GG_2$ form}
We argued in section~\ref{ss:extended} that with $p$ nowhere vanishing vectors $V_A$ we must have
\begin{align}
\Phi &= \Phi^{(3)} + \Phi^{(2)}_{A} \Theta^A  + \ff{1}{3!} \Phi^{(0)}_{ABC} \Theta^{ABC}~.
\end{align}
When $p=3$ we need $\Phi^{(3)} = 0$, since otherwise $\ast_4 \Phi^{(3)}$ will yield an additional vector.  So, we have
\begin{align}
\label{eq:basicPhi}
\Phi &=  \omega_A \Theta^A  + k \Theta^{123}~.
\end{align}
We can take $k\ge 0$, since the sign of $k$ can be changed by redefining $\Theta^A \to -\Theta^A$ and $\omega_A \to -\omega_A$.
This is just a convenient choice of orientation on $X_7$.   

We assume that $\{e^1,e^2,e^3,e^4,\Theta^1,\Theta^2,\Theta^3\}$ is an orthonormal basis for $T^\ast_X$ and check the
compatibility of this with the metric obtained from $\Phi$ via
\begin{align}
g_{ij} = \frac{1}{144} \ep^{klmnpqr} \Phi_{ikl} \Phi_{jmn}\Phi_{pqr}~.
\end{align}
A bit of algebra and~(\ref{eq:basicPhi}) show that
\begin{align}
144 g_{ij} &= 3 \ep^{ABC}\ep^{\alpha\beta\gamma\delta} (\Phi_{i AB}\Phi_{j \alpha\beta} + \Phi_{i \alpha\beta} \Phi_{j AB}) \Phi_{C\gamma\delta} \nonumber\\
&\quad-12 \ep^{ABC} \ep^{\alpha\beta\gamma\delta} \Phi_{i A\alpha}\Phi_{j B \beta} \Phi_{C \gamma\delta} 
+\ep^{ABC}\ep^{\alpha\beta\gamma\delta} \Phi_{i \alpha\beta} \Phi_{j \gamma\delta} \Phi_{ABC}~.
\end{align}
This can be unpacked into various components.  Taking $E_\alpha$ to be the dual vectors to $e^a$, we have the following results:
\begin{align}
g(E_\mu,V_A) & = 0~,  \nonumber\\
g(V_D, V_E) & = \frac{k}{8} \ep^{\alpha\beta\gamma\delta} \Phi_{D\alpha\beta}\Phi_{E\gamma\delta}~, \nonumber\\
g(E_\mu,E_\nu) &   = -\frac{1}{12} \ep^{ABC} \ep^{\alpha\beta\gamma\delta} \Phi_{A \mu\alpha} \Phi_{B \nu \beta} \Phi_{C\gamma\delta}~.
\end{align}

Starting with the general form of $\Phi$, we write $\omega_A = \ff{1}{2} M_{A\alpha\beta} e^{\alpha}\wedge e^{\beta}$, so that $\Phi_{A\alpha\beta} = M_{A\alpha\beta}$.
Finally, setting 
\begin{align}
(\ast M)_{A\alpha\beta} = \ff{1}{2}\ep^{\alpha\beta\gamma\delta} M_{A\gamma\delta}~,
\end{align}
we obtain a simple form for the metric components:
\begin{align}
g(V_D,V_E) &= -\ff{k}{4} \Tr (M_D (\ast M_E))~,&
g(E_\mu,E_\nu) &=-\ff{1}{6} \ep^{ABC} (M_A (\ast M_B) M_C)_{\mu\nu}~.
\end{align}
Since we already verified $g(E_\mu,V_A) = 0$, we now just need to check that
\begin{align}
\label{eq:checkmetric}
\delta_{DE} & = -\ff{k}{4} \Tr (M_D (\ast M_E))~, &
\iden_4 & = -\ff{1}{6} \ep^{ABC} (M_A (\ast M_B) M_C)~.
\end{align}

\subsubsection*{Reduction of parameters by $\SO(4)$ action}
Since~(\ref{eq:checkmetric}) is invariant under $\SO(4)$ rotations $M_A \to R^T M_A R$ we can bring the anti-symmetric matrices $M_A$ to a canonical form.
Without loss of generality we set
\begin{align}
M_1 & = \begin{pmatrix} x_1 \rho & 0 \\ 0 & y_1\rho\end{pmatrix}~,
\end{align}
where $\rho = i\sigma_2$ and $x_1 \ge 0$, $y_1 \ge 0$.\footnote{The $\sigma_i$ are the Pauli matrices.}  This is stabilized by an $\SO(2)\times\SO(2)$ action, which allows us to bring $M_2$
to the form
\begin{align}
M_2 & = \begin{pmatrix} x_2 \rho & P_2 \\ -P_2^T & y_2 \rho \end{pmatrix}~, & P_2 & = \begin{pmatrix} 0 & b_2 \\ c_2 & 0 \end{pmatrix}~,
\end{align}
with $c_2 \ge 0$.  Finally, $M_3$ takes the general form
\begin{align}
M_3 & = \begin{pmatrix} x_3 \rho & P_3 \\ -P_3^T & y_3 \rho 
\end{pmatrix}~,&
P_3 & = \begin{pmatrix} a_3 & b_3 \\c_3 &d_3 \end{pmatrix}~.
\end{align}

\subsubsection*{Solution of the constraints}
We now have a system of 16 equations in~(\ref{eq:checkmetric}) that depend on $13$ parameters:  $12$ of these are in the reduced $M_A$, and $k$ is the last one.
The equations have a unique solution, with
\begin{align}
M_1 & = \begin{pmatrix} 0 &1 & 0 & 0 \\ -1 &0&0 & 0 \\ 0 & 0 & 0 & 1 \\ 0 & 0 & -1 & 0 \end{pmatrix}~, &
M_2 & = \begin{pmatrix} 0 &0 & 0 & 1 \\ 0 &0&1 & 0 \\ 0 & -1& 0 & 1 \\ -1 & 0 & 0 & 0 \end{pmatrix}~, &
M_3 & = \begin{pmatrix} 0 &0 & 1 & 0 \\ 0 &0&0 & -1\\ -1 & 0& 0 & 0 \\ 0 & 1 & 0 & 0 \end{pmatrix}~, &
\end{align}
or in terms of Pauli matrices
\begin{align}
M_1 & = \iden_2 \otimes i\sigma_2~,&
M_2 & = i\sigma_2 \otimes \sigma_1~,&
M_3 & = i\sigma_2 \otimes \sigma_3~.
\end{align}
These satisfy
\begin{align}
\ast M_A &= M_A~,&
M_A M_B &= -\delta_{AB} \iden_4 + \ep_{ABC} M_C~.
\end{align}
Thus,
\begin{align}
\Phi = \omega_A \Theta^A +  \Theta^{123}~,
\end{align}
and the $\omega_A$ are non-degenerate, self-dual, and satisfy $\omega_A\wedge\omega_B = 2 \delta_{AB} e^{1234}$~.

\section{An integral flux for $\cN = 3$ supersymmetry} \label{app:n3flux}
The case of the $\cN=3$ vacuum on $X\times\Xt$  is exotic enough that it is worthwhile to check that it can be obtained by some choice of integral flux and no space-filling M2-branes.\footnote{We thank Dave Morrison for stressing the importance of this point and for discussions regarding the solution presented here and its possible generalizations.}

Let $x = (2A-C)$, and write $f_{\alpha\dot\alpha} = x \phi_{\alpha\dot\alpha}$.  To obtain exactly $\cN=3$ supersymmetry,
we take $x\neq 0$ and the flux must be
\begin{align}
G = -x j_a \jt_{a} + 2x [vE + \vt \Et] + x \omega_\alpha \phi_{\alpha\dot\alpha}\omegat_{\dot\alpha}~.
\end{align}
This flux is integral if and only if
\begin{align}
\frac{1}{2\pi}\left[ -x j_a \jt_{\dot a} + \omega_\alpha f_{\alpha\dot\alpha}\omegat_{\dot \alpha}\right] \in H^4(X\times\Xt,\Z) \qquad
\text{and}\qquad \frac{xv}{\pi} \in \Z~, \qquad \frac{x\vt}{\pi} \in \Z~.
\end{align}
While the implications of the first of these are not immediately obvious, the last two are readily solved:  there are non-zero integers $m$, $\mt$ such that 
\begin{align}
v &= \frac{m\pi}{x}~, &
\vt & = \frac{\mt \pi }{x}~.
\end{align}
The integrated Bianchi identity now becomes
\begin{align}
\label{eq:Bianchileftover}
\frac{m\mt}{2} \left[ 5 +\tr (\phi^T \phi) \right] = 24 - N(M_2)~.
\end{align}
We will now demonstrate that we can choose $m,\mt$ and $\phi$ so that the flux is integral and $N(M_2) = 0$.

Our Ansatz for the flux is motivated by the counting of massless moduli for $\cN = 3$ vacua:  we see that at best, the number of geometric moduli preserved is that of a single K3 geometry, so that it is not unreasonable to tie the geometries of $X$ and $\Xt$ together.  In fact, we will take $X$ and $\Xt$ to be identical.  

Let us explain a little bit more what this means.  We fix an integral basis $\{e^1,e^2,\ldots, e^{22}\}$ for $H^2(X,\Z)$ such that
\begin{align}
e^I e^J = D^{IJ} E~, 
\end{align}
where $D = (-\GE_8)^{\oplus 2} \oplus H^{\oplus 3}$ is the standard metric of signature $(3,19)$.  Since $H^2(X,\Z)$ is unimodular, we have the key fact that $D^{-1}$, with components denoted by $D_{IJ}$ is also an integral matrix.  There is a corresponding set of forms $\et^{\dot I}$ on $\Xt$ that have identical structure.

The $j_a$ and $\omega_\alpha$ can be written in terms of the integral basis:
\begin{align}
j_a &= \cE_{aI} e^I~, &
\omega_\alpha & = \cE_{\alpha I} e^I~.
\end{align}
The coefficients obey
\begin{align}
\label{eq:coeffnorm}
\cE_{aI} D^{IJ} \cE_{bJ}  &= 2 \delta_{ab} v~,&
\cE_{aI} D^{IJ} \cE_{\alpha J} & = 0~, &
\cE_{\alpha I} D^{IJ} \cE_{\beta J} & = -2 \delta_{\alpha\beta} v~.
\end{align}
There is also a useful completeness relation for the vielbeins $\cE$:
\begin{align}
\label{eq:completeness}
\cE_{aI} \cE_{a J} - \cE_{\alpha I} \cE_{\alpha J} = 2 v D_{IJ}~.
\end{align}

We now describe our Ansatz for the flux.
\begin{enumerate}
\item  We assume that $X$ has an integral $-4$ class that is orthogonal to all of the $j_a$.  That is, there exists $\xi \in H^2(X,\Z)$ that is annihilated by the $j_a$ and satisfies $\xi \wedge \xi = \xi \cdot \xi E = - 4 E$.  It is easy to construct smooth K3 geometries with this property at low Picard number.

Without loss of generality we can take $\xi$ to be the direction of one of the anti-self-dual forms.  More precisely, we set
\begin{align}
\omega_1 = \sqrt{\ff{v}{2}} \xi~.
\end{align}

\item 
Once we choose this data for $X$, we use the same $\cE_{a I}$ and $\cE_{\alpha I}$ to prescribe
the $\jt_{\dot a}$ and $\omegat_{\dot\alpha}$, i.e. the geometry of $\Xt$:
\begin{align}
\jt_a &= \cE_{a \dot I} \et^{\dot I}~, &
\omegat_\alpha & = \cE_{\alpha  \dot I} \et^{\dot I}~.
\end{align}
This implies that $v=\vt$, and therefore $m=\mt$ as well; we also have a form $\xit$ as a special $-4$ class on $\Xt$.

\item We take the $\phi_{\alpha\dot\alpha}$ to be diagonal:  $\phi_{\alpha\dot\alpha} = \phi_\alpha \delta_{\alpha\dot\alpha}$.
\end{enumerate}
With these assumptions the flux takes the form
\begin{align}
\frac{G}{2\pi} &= -\frac{x}{2\pi} \left[ j_a \jt_a -\omega_\alpha \omegat_\alpha\right] +\frac{x}{2\pi} (\phi_\alpha-1) \omega_\alpha\omegat_\alpha~ \nonumber\\[2mm]
& = -m D_{I\dot I} e^{I}\et^{\dot I} +\frac{x}{2\pi} (\phi_\alpha-1) \omega_\alpha\omegat_\alpha~,
\end{align}
where in the second line we used the completeness relation~(\ref{eq:completeness}).

The reason this works nicely is that the first term is automatically integral, and we just need to choose the $\phi_\alpha$ appropriately so that the last term is integral as well.  We accomplish this by setting $\phi_\alpha = 1$ for all $\alpha >1$,
so that now
\begin{align}
\frac{G}{2\pi} 
& = -m D_{I\dot I} e^{I}\et^{\dot I} + \frac{m(\phi_1-1)}{4} \xi \xit~.
\end{align}
Choosing $\phi_1 = 5$ leads to an integral flux.  

For this choice of integral flux the Bianchi identity becomes
\begin{align}
\frac{m^2}{2} \left[ 5 +5^2 + 18\right] = 24 - N(M_2)~,
\end{align}
and setting $m=1$, we find the desired $N(M_2) = 0$. 

We have shown that there is a choice of flux that leads to exactly $\cN=3$ supersymmetry without space-filling M2 branes.  The choice leaves many moduli; indeed, the number of massless scalars is smaller than the maximum allowed by just one $\cN=3$ ``hypermultiplet.''  

It is not so easy to generalize this solution.  If one stays with the ``completeness'' relation trick above and simply modifies the $\phi_{\alpha\dot\alpha}$ it is quite likely there are no others with $N(M_2) = 0$.\footnote{There is at least one solution with $N(M_2) \neq 0$; if we set $m =\mt = 1$ and choose $\phi_\alpha = 1$ for all $\alpha$, then we obtain $N(M_2) = 12$.}


\bibliographystyle{./utphys}
\bibliography{./bigref}

\end{document}